# Saikosaponins with similar structures but different mechanisms lead to combined hepatotoxicity

Qianqian Zhang [a, b], Wanqiu Huang [a, b], Yiqiao Gao [a, b], Yingtong Lv [a, b], Wei Zhang [c], Zunjian Zhang [a, b*] and Fengguo Xu [a, b*]

[a] Key Laboratory of Drug Quality Control and Pharmacovigilance (Ministry of Education), China Pharmaceutical University, Nanjing, China

[b] State Key Laboratory of Natural Medicine, China Pharmaceutical University, Nanjing, China

[c] State Key Laboratory for Quality Research in Chinese Medicines, Macau University of Science and Technology, Taipa, Macau, China

[*] **Corresponding Authors**

Zunjian Zhang

Fax/Tel: 86-25-83271454. E-mail: zunjianzhangcpu@hotmail.com.

Fengguo Xu

Fax/Tel: 86-25-83271021. E-mail: fengguoxu@gmail.com.

## ABSTRACT:

*Radix Bupleuri* is a hepatoprotective traditional Chinese medicine (TCM) used for thousands of years in clinical, which was reported to be linked with liver damage. Previous studies have revealed that saikosaponins are the major types of components that contribute to the hepatotoxicity of *Radix Bupleuri*. However the underlying molecular mechanism is far from being understood. In order to clarify whether these structural analogues exert toxicity effects through the same molecular targets, a systematic comparison study was done in this paper. The effects of SSa, b2, c, and d on isolated rat liver mitochondria and human hepatocyte L02 cells were explored, respectively. The collective results indicated that although saikosaponins share the similar structures but they have quite different mechanisms. SSb2 and SSd showed most serious damage on the function of mitochondria and survival rate of cell, respectively. SSb2 could cause mitochondrial permeability transition pore (mPTP) opening and collapse of mitochondrial membrane potential ($\Delta\Psi$m) by impairing the mitochondrial respiratory chain complex III. While SSd destroyed plasma membrane and led to the release of lactate dehydrogenase (LDH) mainly through activating caspase-1. Furthermore, the combine index (CI) demonstrated that the combined hepatotoxicity of SSb2 and SSd could be additive. This finding might yield more in depth understanding of hepatotoxicity of *Radix Bupleuri* possess many different saikosaponins.

**Abbreviations:** TCM, traditional Chinese medicine; SS, saikosaponin; mPTP, mitochondrial permeability transition pore; $\Delta\Psi$m, mitochondrial membrane potential; CI, combine index; LDH, lactate dehydrogenase; AA, antimycin A; Rot, rotenone; DCFH-DA, 2′,7′-dichlorofluorescein diacetate; Rh123, rhodamine 123; z-VAD-fmk, N-benzyloxycarbonyl-Val-Ala-Asp-fluoromethylketone; GSH, glutathione; ROS, reactive oxygen species.

# 1. Introduction

*Radix Bupleuri* is one of the most important TCMs and has been used in China for thousands of years [1]. In clinical, many herbal formulas containing *Radix Bupleuri* have been shown to be effective in treating hepatic injury syndromes, for instance Sho-saiko-to (TJ-9) [2], Inchinko-to (TJ-135) [3], Han-Dan-Gan-Le [4], Shugan-Huayu powder (SHP) [5]. Simultaneously, increasing clinical and experimental evidences concerning liver injuries were found along with the wide usage of *Radix Bupleuri* [6, 7]. Itoh.S *et al.* reported that patients treated with TJ-9 exhibited acute drug-induced liver injury during the latent period of one and a half to three months [8]. Lee *et al.* observed herbs containing more than 19 grams of *Radix Bupleuri* in HBV-infected patients would increase the risks of liver injury [9]. Furthermore, extracts of *Radix Bupleuri* could cause hepatic injury for oxidative damage [10]. Hence, *Radix Bupleuri* may increase liver injury when used in the treatment of liver diseases.

Pharmacology and chemical composition of *Radix Bupleuri* had been studied to find which component causes liver toxicity. Saikosaponins (major series: SSa, b2, c, and d) were believed to be the main compounds responsible for liver toxicity of *Radix Bupleuri* [10]. Saikosaponins induced the hepatoxicity by causing liver cell damage and necrosis after administrating continuously to rats for 15 days [10]. Another study proved that SSd disrupted platelet-derived growth factor-β receptor/p38 pathway and thus led to mitochondrial apoptosis in L02 cells [11]. Further investigation demonstrated that SSd activated Fas, caspase-8, and Bid, contributing to cytochrome c release and caspase-3 activation in vitro [12]. Additionally, SSa resulted in plasma membrane injury and finally caused cytotoxicity by reducing activity of SOD [10]. However, the underlying hepatotoxicity mechanism of saikosaponins is far from clearly elucidated.

Saikosaponins are pentacyclic triterpene compounds (Fig. 1). Despite there are slight structural differences among them, their pharmacological activities, mechanisms and applications vary dramatically [13, 14]. Therefore, in this study we comparatively studied the toxicity of SSa, b2, c, and d on isolated rat liver mitochondria and human hepatocyte L02 cells, as well as the combined toxicity of SSb2 and SSd. SSb2 and SSd showed most serious damage on mitochondrial function and cell viability, respectively, and the combined hepatotoxicity of the two components was found to be additive. This research might provide a new and comprehensive insight into the hepatotoxicity of different saikosaponins.

## 2. Material and methods

### 2.1. Materials

Saikosaponin standards (SSa, SSb2, SSc, SSd, purity > 98%) were purchased from Sichuan Victory Biological Technology Co., Ltd. (Chengdu, China). Sucrose, mannitol, Hepes, Tris-HCl, KCl, $CaCl_2$, EGTA were all purchased from Nanjing SunShine Biotechnology Co., Ltd. (Nanjing, China). Cyclosporin A (CsA), carbonyl cyanide 3-chlorophenylhydrazone (CCCP), antimycin A (AA), rotenone (Rot), 2′,7′-dichlorofluorescein diacetate (DCFH-DA), rhodamine 123 (Rh123), sodium succinate, pyruvate, malate, glutathione (GSH), DMSO were all purchased from Sigma-Aldrich (St. Louis, MO, USA). N-benzyloxycarbonyl-Val-Ala-Asp-fluoromethylketone (z-VAD-fmk) was purchased from APExBIO (Houston, USA). RPMI-1640 medium was purchased from Gibco (Grand Island, NY, USA). Fetal bovine serum (FBS) was purchased from Sijiqing Biological Engineering Materials Co., Ltd. (Hangzhou, China).

### 2.2. Preparation of mitochondria

Mitochondria were isolated from the rat liver in a cold isolation medium containing 220 mM mannitol, 70 mM sucrose, 2 mM Hepes and 1 mM EGTA (PH=7.4) by differential centrifugation [15]. The isolated mitochondria were in a pellet form and suspended in incubation medium containing 220 mM mannitol, 70 mM sucrose and 2 mM Hepes (PH=7.4) at 4 ℃. Mitochondrial protein concentration was determined by the Bradford protein assay kit (Beyotime Institute of Biotechnology, Shanghai, China) according to the protocols..

### 2.3. Determination of mitochondrial swelling and ΔΨm

Mitochondrial swelling, another subsequence of mPTP, and ΔΨm were evaluated as described previously [16]. $CaCl_2$ (50 μM) and CCCP (50 μM) were used as the 100% baseline for swelling and ΔΨm, respectively. The induction of mitochondrial swelling was monitored at 540 nm absorbance (A540) by using Tecan Infinite 200 Pro (Tecan Group Ltd, Mannedorf, Switzerland). A decrease in the absorbance indicates an increase in mitochondrial swelling and the opening of mPTP. Mitochondrial uptake of the cationic fluorescent dye, Rh123, has been used for the estimation of $\Delta\Psi m$. $\Delta\Psi m$ was monitored at the excitation and emission wavelength of 490 nm and 535 nm, respectively, using POLARstar Omega Plate Reader Spectrophotometer (BMG LABTECH, Ortenberg, Germany). An increase in the fluorescence indicates the release of Rh123 and collapse of $\Delta\Psi m$.

## 2.4. Quantification of mitochondrial reactive oxygen species level

The mitochondrial reactive oxygen species (ROS) level was measured using DCFH-DA fluorescence probe as described [17] which could permeate through mitochondrial inner membrane and be trapped in mitochondrial matrix, where it is reduced to lipophobic DCFH and then oxidized to fluorescent DCF by ROS. The production of ROS measured through POLARstar Omega Plate Reader Spectrophotometer equipped with the excitation wavelength of 488 nm and emission wavelength 526 nm. Mitochondrial complex I inhibitor Rot and complex III inhibitor AA were added to determine which complex was damaged.

## 2.5. Cell culture

L02 cells were purchased from the Cell Bank of Chinese Academy of Sciences (Shanghai, China) and were cultured in RPMI-1640 medium supplemented with 10% heat-inactivated FBS, and grown in a 95% air and 5% $CO_2$ humidified atmosphere at 37 ℃.

## 2.6. Cell viability assay

L02 cells were seeded in 96-well plates and cultured in RPMI-1640 supplemented with 10% FBS for 24 h, and then treated with different concentration of saikosaponins at the indicated concentrations for 24 h. After treatment, CCK-8 (Beyotime Institute of Biotechnology, Shanghai, China) was added into each well and incubated at 37 ℃ for 2 h, and then the 450 nm absorbance of each well was by Tecan Infinite 200 Pro microplate reader.

## 2.7. Flow cytometric analysis of apoptosis and necrosis

Apoptosis, or programmed cell death was performed in a Muse™ cell analyzer (Merck Millipore, Billerica, MA, USA) utilizing a Muse annexin V & dead cell kit (MCH100105; Merck Millipore). The assay utilizes phycoerythrin-labeled annexin V to detect phosphatidylserine (PS) on the external membrane of apoptotic cells. In addition, the kit contains 7-aminoactinomycin D (7-AAD), the DNA dye, to detect the cells without plasma membrane structural integrity. Four populations of cells can be distinguished in this assay: nonapoptotic cells, early apoptotic cells, late-stage apoptotic and dead cells, necrotic nuclear debris. The assay was performed according to the manufacturer's protocols. L02 cells were seeded in 6-well plates and cultured in RPMI-1640 supplemented with 10% FBS for 24 h, and then treated with different substrates at indicated concentrations. Both adherent and floating cells were harvested and washed once with PBS. 100 μL of annexin V & dead reagent and 100 μL of a single cell suspension were mixed in a microtube and incubated for 20 min at room temperature in the dark. 5,000 cell events were collected for each sample, and the percentages of apoptotic and necrotic cells were determined by the Muse™ cell analyzer.

## 2.8. LDH release assay

Cells treatment with the cytotoxic compound may undergo necrosis, in which they lose membrane integrity. LDH is a stable enzyme exists in nearly all living cells, which is rapidly released into the cell culture supernatant when the plasma membrane damages. Hence the quantification of plasma membrane damage can be detected by LDH in the cell culture supernatant. LDH activities in culture medium were determined using a LDH release assay kit (Beyotime Institute of Biotechnology, Shanghai, China) according to the protocols. L02 cells were seeded in 96-well plates and incubated in RPMI-1640 supplemented with 10% FBS for 24 h, and then were treated with different substrates at indicated concentrations. The spectrophotometric absorbance at 490 nm was determined by Tecan Infinite 200 Pro.

## 2.9. Measurement of caspase-1 activity

Caspase-1 activity was measured using a Caspase-1 assay kit (Beyotime Institute of Biotechnology, Shanghai, China). L02 cells were seeded in 6-well plates and cultured in RPMI-1640 supplemented with 10% FBS for 24 h, and then treated with SSd for 24 h. Cells were harvested and centrifuged at 600 g, 4 ℃ for 10 min, and the pellet containing $2 \times 10^6$ cells was incubated in 100 μL ice-cold lysis buffer for 15 min and then centrifuged at 16000 g, 4 ℃ for 10 min. Supernatants were retrieved and aliquots corresponding to protein at the concentration of 3 mg/ml for each assay. 40 μL of caspase-1 buffer and 10 μL of the 2 mM YVAD-pNA substrate were added to each sample. The samples were then incubated at 37 ℃. The absorbance was recorded at 405 nm using Tecan Infinite 200 Pro. The activity of caspase substrates was expressed as a percentage of enzyme activity compared with the controls.

**2.10. Combined toxicity of SSb2 and SSd**

To determine the combined toxicity of SSb2 and SSd, CI, a commonly used evaluation index, was calculated. The combined toxicity of SSb2 and SSd was calculated by the following equation:

$$CI = \frac{(1 - V_{b2} \times V_d)}{(1 - V_{com})}$$

$V_{com}$ is cell viability of L02 cells after treated by SSb2 and SSd, $V_{b2}$ is cell viability of L02 cells after treated by SSb2, $V_d$ is cell viability of L02 cells after treated by SSd. CI<0.9 indicates a synergistic effect, 0.9<CI<1.1 indicates an additive effect, and CI>1.1 indicates an antagonistic effect [18, 19].

**2.11. Statistical analysis**

Each experiment in the present study was repeated 3 times independently. Statistical analysis for comparison of 2 groups was performed using the unpaired Student's t test, utilizing the Prism software (version 6.0, GraphPad).

## 3. Results

**3.1. SSb2 significantly damages mitochondrial function**

Mitochondrial swelling and ΔΨm, as mitochondrial functional parameters, were used to evaluate the drug toxicity in mitochondrial dysfunction. The induction of mitochondrial swelling by different concentrations of saikosaponins (6.25, 12.5, 25, 50, 100 μM) was shown in Fig. 2A-E. The result indicated that SSb2 could significantly induce mPTP opening in a concentration-and time-dependent manner compared with the other three saikosaponins (SSa, c and d). Meanwhile, different concentrations of saikosaponins (6.25, 12.5, 25, 50, 100 μM) decreased ΔΨ m in different degree, which was demonstrated in Fig. 2F, SSb2 destroyed ΔΨ most seriously compared with SSa, c and d. Our result showed that SSb2 significantly damage the function of mitochondria.

**3.2. Impairment of complex III by SSb2 leads to increased ROS production**

Previous studies have demonstrated the close relation between mPTP opening and ROS increasing [20, 21]. To examine whether SSb2 induced mPTP opening was associated with

ROS production, mitochondrial ROS level was measured by fluorescence dye DCFH-DA. As shown in Fig. 3A, SSb2 induced the generation of mitochondrial ROS in succinate buffer, implying that SSb2 increase mPTP opening might through promoting ROS level. Meanwhile, mitochondrial swelling could be effectively prevented by GSH (Fig. 3B), an antioxidant, revealing that SSb2-induced mPTP opening has been caused by the increase of ROS production.

Studies proved that both complexes I and III of mitochondrial respiratory chain are involved in ROS formation [22, 23]. To further explore the possible target of SSb2 on mitochondria respiratory chain, complex I inhibitor Rot and complex III inhibitor AA were added to test how SSb2 influence the generation of mitochondrial ROS in malate or succinate buffer, respectively. As shown in Fig. 3C-D, AA significantly enhanced ROS generation induced by SSb2 both in malate and succinate buffer, which suggested the impairment of complex III activity is a potential source of SSb2 contributed to ROS formation in isolated mitochondria.

### 3.3. SSd significantly induce cell death

More and more studies showed the opening of mPTP plays an important role in both necrosis and apoptosis process [24, 25]. Since SSb2 markedly affected mPTP compared with SSa, c and d, could SSb2 damage the survival rate of cells mostly? To examine whether SSb2 could induce cell death in the most degree, cell viability assay, CCK-8 test, Hoechst 33258 fluorescence staining and Annexin-V/7-AAD test were conducted after the addition of different saikosaponins (SSa, b2, c and d). As displayed in Fig. 4A-C, while, SSd could significantly induce cell death instead of SSb2.

### 3.4. SSd damages the plasma membrane by activating caspase-1

As demonstrated in Fig. 4C, SSd could significantly destroy plasma membrane compared with the other three saikosaponins (SSa, b2 and c). Additionally, SSd could destroy plasma membrane and increase the release of LDH in a concentration-dependent (Fig. 5A). Recent studies showed that activated caspase-1 could lead to the impairment of plasma membrane [26, 27]. To investigate whether caspase-1 is involved in the damage of plasma membrane induced by SSd, the activity of caspase-1 was measured. Caspase-1 was markedly activated at 1.76 times the intensity of the activation in the control cells after treatment with

17.5 μM of SSd (Fig. 5B). Furthermore, live cell rate increased and LDH release decreased significantly after the pre-treated with z -VAD-fmk (20 μM), a commonly used caspase inhibitor, for 2 h before the addition of SSd (Fig. 5C-D). This result indicated that SSd-induced the damage of plasma membrane resulted from the activation of caspase-1, which could be significantly prevented by the caspase inhibitor.

### 3.5. Combined toxicity of SSb2 and SSd origins from additive effect

Liver injuries caused by *Radix Bupleuri* might result from a combined hepatotoxicity of multi-components. SSb2 and SSd are the representative type I saikosaponin and type II saikosaponin in *Radix Bupleuri*, respectively, which could damage mitochondrial function and cell viability with different mechanisms. Hence, to explore the combined toxicity of SSb2 and SSd may get a new insight into the liver toxicity mechanism caused by *Radix Bupleuri*. CI was calculated and shown in Fig. 6A, which indicated the combined toxicity of SSb2 and SSd was an additive effect. Moreover, the combined toxicity of SSb2 and SSd is significantly serious than the single saikosaponin treatment tested by Annexin-V/7-AAD assay, as displayed in Fig. 6B.

## 4. Discussion

Isolated liver mitochondria and L02 cells were used to examine the toxicity of SSa, b2, c, and d. As a result, SSb2 and SSd were found to be the most toxic components in the two different models, respectively. And then the toxicity mechanisms and combined hepatotoxicity of these two saikosaponins were studied. Our inferred comprehensive hepatoxicity mechanisms of SSb2 and SSd were summarized in Fig. 7.

Saikosaponins (SSa, b2, c, d) are the major hepatotoxicity components of *Radix Bupleuri* [10]. However which saikosaponin is dominant in liver injury caused by *Radix Bupleuri* is not well known, as well as its underlying molecular mechanism. The in-vitro models are widely used in rapid and effective toxicity assessing and investigating the hepatotoxicity mechanism of targeted compounds, for instance isolated liver mitochondria and L02 cells models [16, 28], which can provide a new understanding of hepatotoxicity mechanism of saikosaponins.

Mitochondria are primary sites for a number of vital metabolic processes and critical targets for many toxins as well as origins of several diseases, which play an important role in the cell death as well as many physiological and pathological phenomena such as participation in ROS formation, liver injury, obesity, inflammation and aging [29-31]. Many studies confirmed that there are many factors contributing to the opening of mPTP, including the increased Pi, $Ca^{2+}$, ROS, RNS, $NAD(P)^+$, GSSG and decreased ATP [32, 33]. Base on our results, SSb2 could cause mPTP opening and lead to collapse of ΔΨm by increasing ROS production. The further investigation indicated that the impairment of complex III activity is a potential source of SSb2 induced ROS formation in isolated liver mitochondria (Fig. 3). It has been revealed that complex III is a major site for superoxide anion generation, an important source of oxygen radicals, in mitochondria [17, 20]. The impairment of complex III by SSb2 may enhance the leakage of electrons from the electron transport chain, generating more oxide radicals, which is responsible for SSb2 induced mitochondrial ROS production that ultimately causes the opening of mPTP and mitochondrial dysfunction.

On the other hand, SSd was found to be the most toxicant saikosaponin for L02 cells according to our results. Recent studies proved that activated caspase-1 can specifically cleave the linker between the carboxy-terminal gasdermin-C and amino-terminal gasdermin-N domains in gasdermin D (GSDMD), which lead to the impairment of cell membrane [26, 27]. Based on our data, caspase-1 was activated after SSd treated (Fig. 5B). Meanwhile, z-VAD-fmk could significantly inhibit the plasma membrane damage caused by SSd (Fig. 5C-D). Therefore, SSd damaged plasma membrane and led to L02 cells death might through activating caspase-1.

SSb2 could impair the function of mitochondria mostly compared with SSa, c and d, while SSd showed most serious damage on cell viability. This result showed that the toxicity targets of SSb2 and SSd were different, which might be attributed to the structural differences of the two compounds. Saikosaponins are oleanane type triterpenoid saponins and divided into seven types according to different aglycones[10]. SSd is epoxy-ether saikosaponin (type I), while SSb2 is heterocyclic diene saikosaponin (type II) [34]. SSb2 gets double bond and hydroxyl group via cleavage of epoxy ether bond in SSd [35]. Compared with SSd, SSb2 is stronger nucleophilic. So it is reasonable to consider that SSb2 has a more destructive effect on the mitochondrial respiratory chain. Meanwhile, SSb2 showed lighter damage on L02 cells than SSd, which might be due to lower amounts of SSb2 entering into the cells.

Hepatotoxicity of *Radix Bupleuri* is derived from its combinations of toxic components. Hence, combinations of saikosaponins can also potentially clarify mechanisms of its herbal toxicity. Based on CI, the combined toxicity of SSb2 and SSd could be an additive effect (Fig. 6). Previous studies revealed that SSd could transform into SSb2 under the condition of controlling temperature and acid [35, 36]. So, the combined toxicity of SSb2 and SSd would be reduced after some SSd transformed into SSb2, since the toxicity on L02 cells of SSd is more serious than SSb2 when in the same quantity. Meanwhile, the efficacy of *Radix Bupleuri* could be interrupted along with SSd transforming into SSb2, because SSd is the main reported bioactive ingredient in *Radix Bupleuri*, possesses anti-inflammatory, antitumor, immunoregulation and other pharmacological activities [37-40]. Therefore, in order to decrease the hepatotoxicity and ensure the efficacy of *Radix Bupleuri*, the amount of SSb2 and SSd should be considered.

## 5. Conclusion

Our results clarified that saikosaponins with slight structural differences exert toxicity effects through the different molecular targets. SSb2 induced ROS levels by impairing the complex III of mitochondrial respiratory chain, while SSd damaged plasma membrane through activating caspase-1. And the combined hepatotoxicity of these two saikosaponins with similar structures components was additive.

## Acknowledgements

This work was supported by the NSFC (No. 81573626), Macao Science and Technology Development Fund (FDCT, No.006/2015/A1), the Program for Jiangsu province Innovative Research Team, the Program for New Century Excellent Talents in University (No. NCET-13-1036), a project funded by the Priority Academic Program Development of Jiangsu Higher Education Institutions (PAPD) and the Open Project Program of Guangxi Key Laboratory of Traditional Chinese Medicine Quality Standards.

**Figure captions**

**Fig. 1.** Chemical structures of (A) SSa, (B) SSb2, (C) SSc and (D) SSd.

**Fig. 2.** SSb2 significantly damages mitochondrial function. (A-E) Mitochondrial swelling. Mitochondria isolated from rat liver were treated with different concentrations of saikosaponins (A: 6.25 μM, B: 12.5 μM, C: 25 μM, D: 50 μM, E: 100 μM). (F) Loss of ΔΨm. Collapse of ΔΨm induced by different concentrations of saikosaponins (0, 6.25, 12.5, 25, 50, 100 μM).

**Fig. 3.** Impairment of complex III by SSb2 leads to increased ROS production. (A) ROS formation after addition of various concentrations of SSb2 (0, 6.25, 12.5, 25, 50, 100 μM). (B) Mitochondrial swelling. Mitochondria were treated with 50 μM SSb2 and 200 μM GSH+SSb2 (GSH+SSb2). (C-D) Effects of respiratory complexes I and III substrates and inhibitors on SSb2-induced ROS formation in isolated rat liver mitochondria, *$P<0.05$, **$P<0.01$. (C) ROS formation was measured in the presence of: 5 mM succinate alone (Suc), succinate+50μM SSb2 (Suc+SSb2), succinate+SSb2+2 μM antimycin A (Suc+SSb2+AA), succinate +SSb2 + 2 μM rotenone (Suc+SSb2+Rot). (D) ROS formation was measured in the presence of: 5 mM pyruvate+10 mM malate alone (Pyr/mal), pyruvate/malate+50μM SSb2

(Pyr/mal+SSb2), pyruvate/malate+SSb2+ 2 μM antimycin A (pyr/mal+SSb2+AA), pyruvate/malate + SSb2+2 μM rotenone (Pyr/mal+SSb2+Rot).

**Fig. 4.** SSd significantly induce cell death. (A) CCK-8 assay. L02 cells were treated with different concentrations of saikosaponins for 24 h (10, 15, 25, 50,100 μM). (B) Hoechst 33258 fluorescence staining. L02 cells were treated with the indicated concentrations of saikosaponins (25 μM) for 24 h. (C) Flow cytometric analysis of apoptosis. L02 cells were treated with the indicated concentrations of saikosaponins (25 μM) for 24 h.

**Fig. 5.** SSd damages plasma membrane by activating caspase-1. (A) LDH release assay. L02 cells were treated with different concentrations of SSd for 24 h (0, 10, 12.5, 15, 17.5, 20 μM). (B) Measurement of caspase-1 activity. L02 cells were treated with SSd (17.5 μM) for 24 h, *P<0.05; **P<0.01. (C) LDH release assay. L02 cells were pertreated with z-VAD-fmk (2 h, 20 μM ) and then treated with SSd for 24 h (17.5 μM), SSd vs. Control, *P <0.05, **P <0.01; z-VAD-fmk+SSd vs. SSd, #P<0.05, ##P<0.01. (D) Flow cytometric analysis of apoptosis. L02 cells were pertreated with z-VAD-fmk (2 h, 20 μM ) and then treated with SSd (15, 17.5, 20 μM) for 24 h.

**Fig. 6.** Measurement of combine toxicity of SSb2 and SSd. (A) Combine index. Combined concentration of SSb2 (6.25, 12.5, 25, 50, 100 μM) plus SSd (10, 12.5, 15μM). (B) Flow cytometric analysis of apoptosis. L02 cells were treated with SSb2 (50 μM), SSd (12.5 μM) and SSb2 (50 μM) + SSd (12.5 μM).

**Fig. 7.** Hepatoxicity mechanisms combined with SSb2 and SSd. RET, reversed electron transfer; FET, forwarded electron transfer.

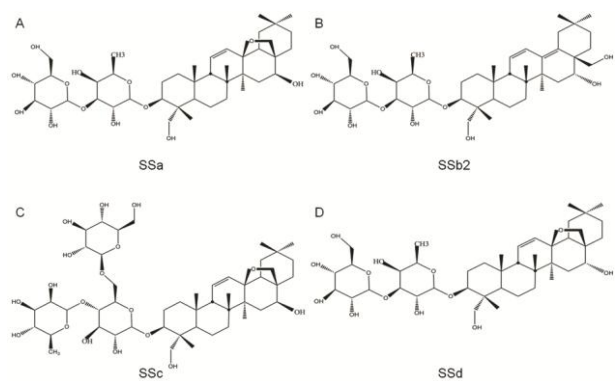

**Fig. 1.** Chemical structures of (A) SSa, (B) SSb2, (C) SSc and (D) SSd.

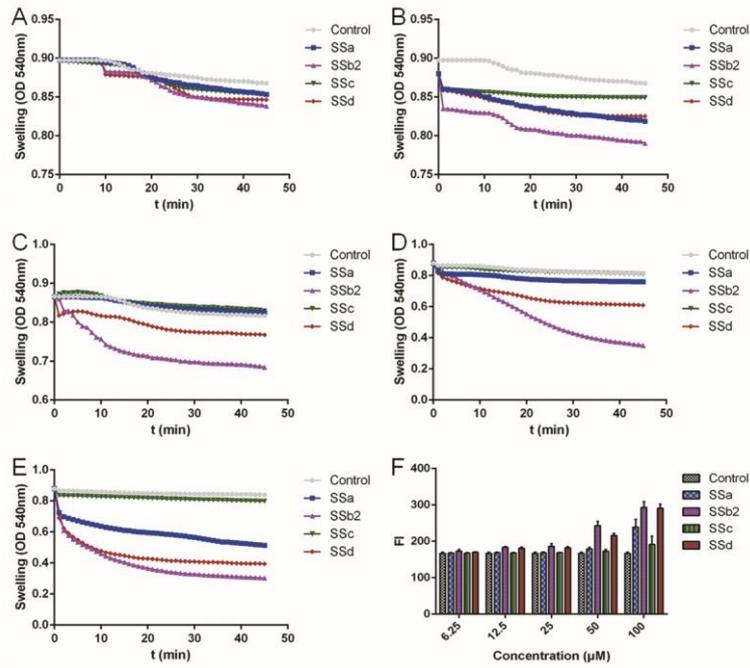

**Fig. 2.** SSb2 significantly damages mitochondrial function. (A-E) Mitochondrial swelling. Mitochondria isolated from rat liver were treated with different concentrations of saikosaponins (A: 6.25 μM, B: 12.5 μM, C: 25 μM, D: 50 μM, E: 100 μM). (F) Loss of ΔΨm. Collapse of ΔΨm induced by different concentrations of saikosaponins (0, 6.25, 12.5, 25, 50, 100 μM).

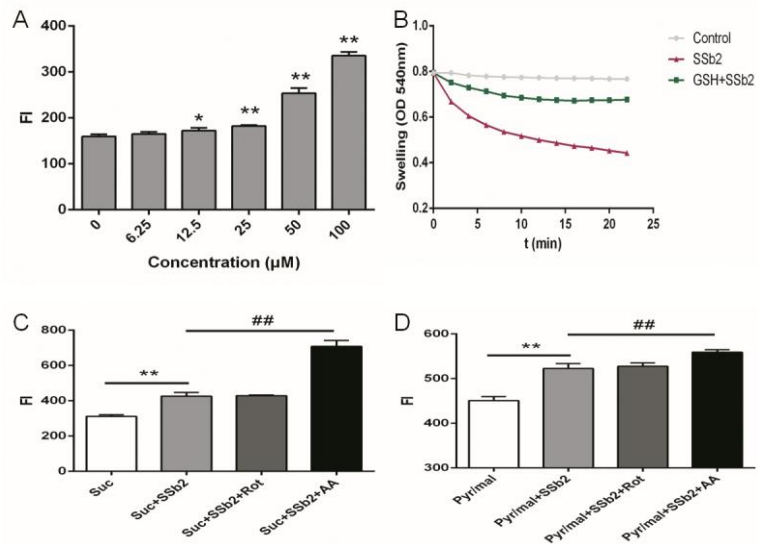

**Fig. 3.** Impairment of complex III by SSb2 leads to increased ROS production. (A) ROS formation after addition of various concentrations of SSb2 (0, 6.25, 12.5, 25, 50, 100 μM). (B) Mitochondrial swelling. Mitochondria were treated with 50 μM SSb2 and 200 μM GSH+SSb2 (GSH+SSb2). (C-D) Effects of respiratory complexes I and III substrates and inhibitors on SSb2-induced ROS formation in isolated rat liver mitochondria, *$P<0.05$, **$P<0.01$. (C) ROS formation was measured in the presence of: 5 mM succinate alone (Suc), succinate+50μM SSb2 (Suc+SSb2), succinate+SSb2+2 μM antimycin A (Suc+SSb2+AA), succinate +SSb2 + 2 μM rotenone (Suc+SSb2+Rot). (D) ROS formation was measured in the presence of: 5 mM pyruvate+10 mM malate alone (Pyr/mal), pyruvate/malate+50μM SSb2 (Pyr/mal+SSb2), pyruvate/malate+SSb2+ 2 μM antimycin A (pyr/mal+SSb2+AA), pyruvate/malate + SSb2+2 μM rotenone (Pyr/mal+SSb2+Rot).

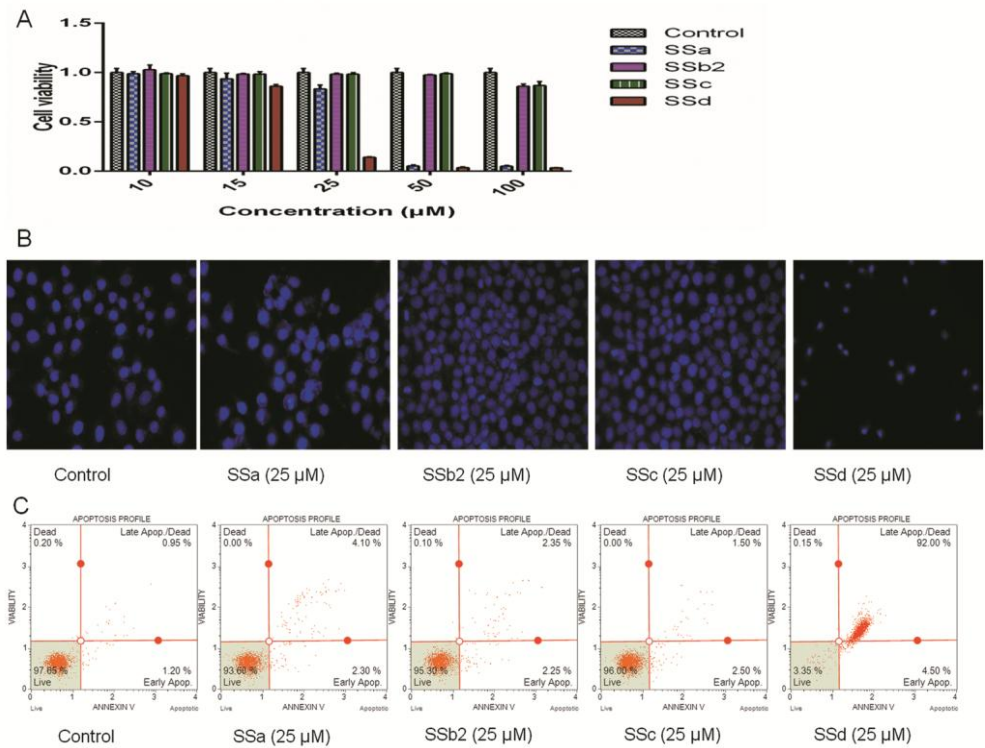

**Fig. 4.** SSd significantly induce cell death. (A) CCK-8 assay. L02 cells were treated with different concentrations of saikosaponins for 24 h (10, 15, 25, 50,100 μM). (B) Hoechst 33258 fluorescence staining. L02 cells were treated with the indicated concentrations of saikosaponins (25 μM) for 24 h. (C) Flow cytometric analysis of apoptosis. L02 cells were treated with the indicated concentrations of saikosaponins (25 μM) for 24 h.

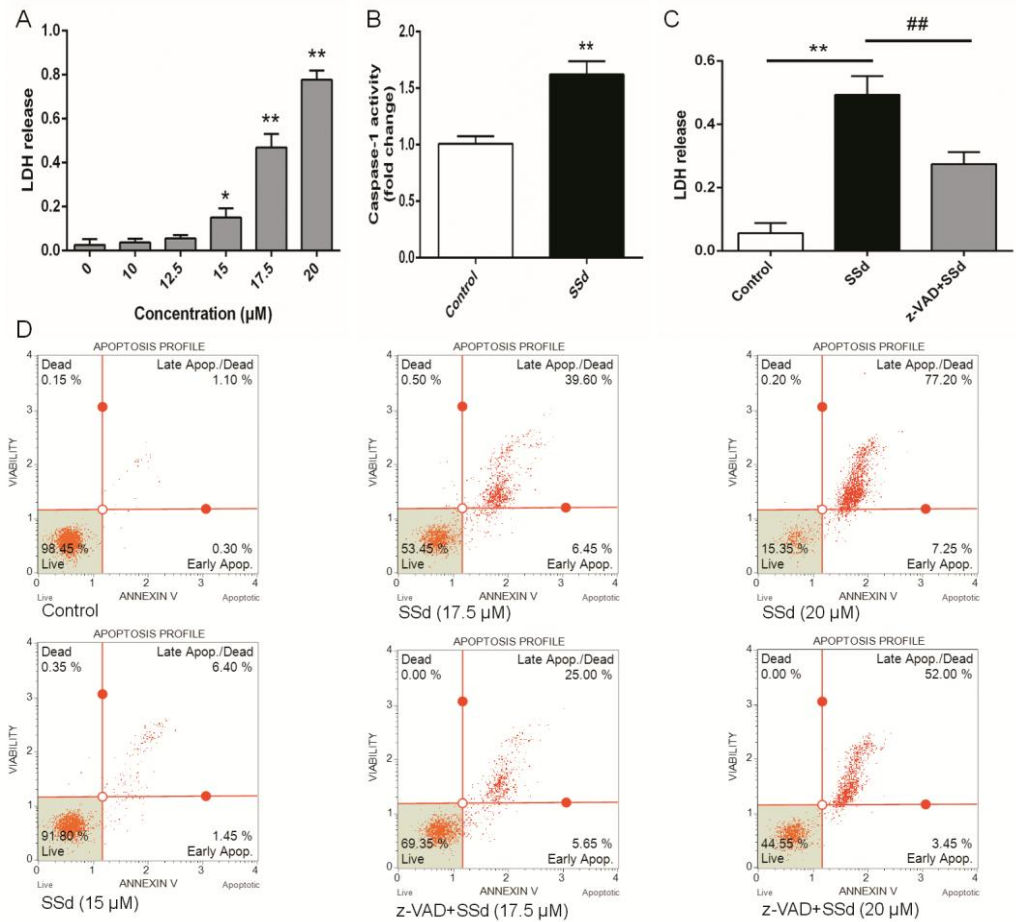

**Fig. 5.** SSd damages plasma membrane by activating caspase-1. (A) LDH release assay. L02 cells were treated with different concentrations of SSd for 24 h (0, 10, 12.5, 15, 17.5, 20 μM). (B) Measurement of caspase-1 activity. L02 cells were treated with SSd (17.5 μM) for 24 h, *P<0.05; **P<0.01. (C) LDH release assay. L02 cells were pertreated with z-VAD-fmk (2 h, 20 μM ) and then treated with SSd for 24 h (17.5 μM), SSd *vs*. Control, *P <0.05, **P <0.01; z-VAD-fmk+SSd *vs*. SSd, #P<0.05, ##P<0.01. (D) Flow cytometric analysis of apoptosis. L02 cells were pertreated with z-VAD-fmk (2 h, 20 μM ) and then treated with SSd (15, 17.5, 20 μM) for 24 h.

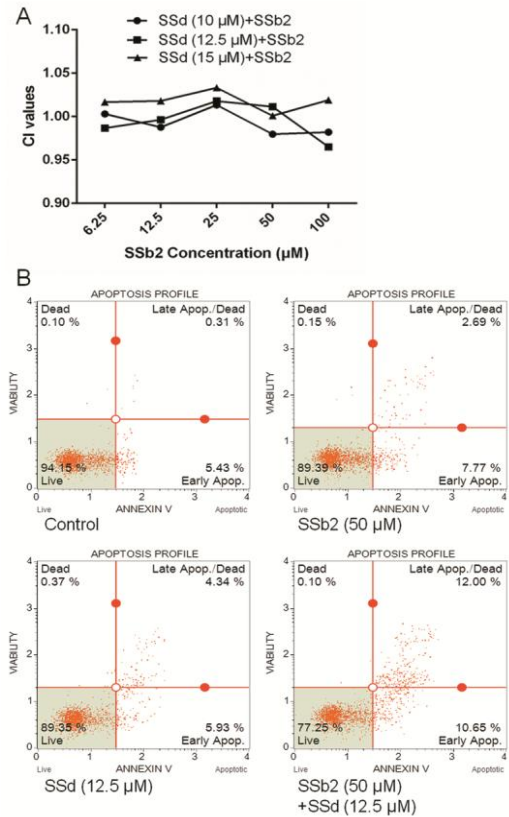

**Fig. 6.** Measurement of combine toxicity of SSb2 and SSd. (A) Combine index. Combined concentration of SSb2 (6.25, 12.5, 25, 50, 100 μM) plus SSd (10, 12.5, 15μM). (B) Flow cytometric analysis of apoptosis. L02 cells were treated with SSb2 (50 μM), SSd (12.5 μM) and SSb2 (50 μM) + SSd (12.5 μM).

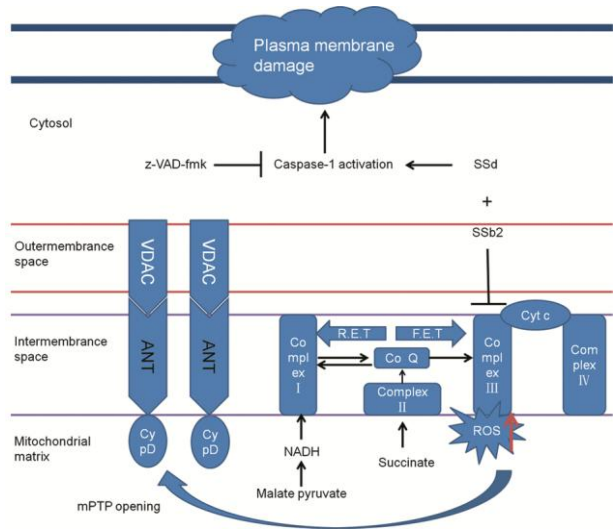

**Fig. 7.** Hepatoxicity mechanisms combined with SSb2 and SSd. RET, reversed electron transfer; FET, forwarded electron transfer.